\documentclass[amsmath,amssymb,aps,pra,twocolumn]{revtex4-1}

\usepackage{graphicx}
\usepackage{dcolumn}
\usepackage{bm}
\usepackage{longtable}
\usepackage{slashed}
\usepackage{epsfig}
\usepackage{dsfont}
\usepackage{physics}
\usepackage{textcomp}

\begin{document}

\title{Semiclassical magnetotransport including the effects of the Berry curvature and Lorentz force}

\author{Seungchan Woo$^{1, 2}$}
\thanks{These authors contributed equally to this work.}
\author{Brett Min$^{3, 2}$}
\thanks{These authors contributed equally to this work.}
\author{Hongki Min$^{1, 2}$}
\email{hmin@snu.ac.kr}
\affiliation{$^1$ Department of Physics and Astronomy, Seoul National University, Seoul 08826, Korea}
\affiliation{$^2$ Center for Theoretical Physics (CTP), Seoul National University, Seoul 08826, Korea}
\affiliation{$^3$ Department of Physics, McGill University, Montr\'{e}al, Qu\'{e}bec H3A 2T8, Canada}
\date{\today}

\begin{abstract}

In topological semimetals and insulators, negative longitudinal magnetoresistance and angle-dependent planar Hall effect have been reported arising from the Berry curvature. Using the Boltzmann transport theory, we present a closed-form expression for the nonequilibrium distribution function which includes both the effects of the Berry curvature and Lorentz force. Using this formulation, we obtain analytical expressions for conductivity and resistivity tensors in Weyl semimetals demonstrating a characteristic field dependence arising from the competition between the two effects.

\end{abstract}

\maketitle

\section{Introduction}

In topological materials with a nonvanishing Berry curvature, the positive magnetoconductance has been observed experimentally in the presence of parallel electric and magnetic fields \cite{KIM2013, HUANG2015, XIONG2015, LI2015, LI2016a, ZHANG2016, LI2016b, ZHANG2016, ZHANG2017, WANG2012, HE2013, WANG2015, WIEDMANN2016, ARNOLD2016, BREUNIG2017, ASSAF2017}. This is a unique feature to topological materials that does not occur in conventional magnetotransport which only takes classical Lorentz force effect into account. 

The prevailing explanation for the positive longitudinal magnetoconductance is the so-called chiral anomaly. In 1983, Nielsen and Ninomiya suggested the chiral anomaly in Weyl fermions under a strong magnetic field regime where the chiral zeroth Landau level creates a one-dimensional conducting channel that pumps electrons from one Weyl node to another \cite{NIELSEN1983}. In 2013, Son and Spivak discussed chiral anomaly in Weyl semimetals under weak external magnetic field using the semi-classical Boltzmann approach \cite{SON2013}. They argued the positive magnetoconductivity that scales quadratically in magnetic field is due to topological charge pumping. One can expect possible detection of chiral anomaly between the valleys, given that the intervalley scattering is negligible compared to the intravalley scattering \cite{SEKINE2017}. However, chiral anomaly cannot be responsible for observed positive magnetoconductivity in topological insulators (TIs) where chiral charges are not well defined \cite{WANG2012, HE2013, WANG2015, WIEDMANN2016, ARNOLD2016, BREUNIG2017, ASSAF2017}. In topological materials such as TIs \cite{DAI2017} and Weyl semimetals (WSMs) \cite{KIM2014}, it is suggested that the anomalous velocity induced by the non-trivial Berry curvature alone can generate an additional contribution to the conductivity that grows with the magnetic field. 

To understand magnetotransport properties quantitatively, it is important to consider both the effects of the anomalous velocity due to the non-trivial Berry curvature and the classical Lorentz force effect. Most of the previous studies \cite{KIM2014, DAI2017, NANDY2017} have focused on the Berry curvature effect, while only a few took the Lorentz force into considerations in describing magnetotransport behaviors \cite{IMRAN2018, JOHANSSON2019}. In this paper, we revisit semiclassical treatment of magnetotransport in topological materials to shed more light on the origin of the observed positive magnetoconductivity. We present a general semiclassical formula for conductivity which fully incorporates the Berry curvature and the Lorentz force. From the Boltzmann transport equation, we obtain a closed-form expression for the nonequilibrium distribution function by solving the corresponding self-consistent equation. We then apply our formula to WSMs and express the magnetoconductivity in terms of dimensionless parameters characterizing magnetic fields associated with the Lorentz force and the Berry curvature, respectively.


\section{Semiclassical Boltzmann magnetotransport theory for topological materials}


The semiclassical Boltzmann transport equation governs the time evolution of a non-equilibrium distribution function $f=f(\mathbf{r},\mathbf{k},t)$ at position $\mathbf{r}$ and momentum $\mathbf{k}$. It states that the time evolution of $f$ equals the probability rate of electrons being scattered in and out of the distribution, called the collision term $\Big(\frac{\partial f}{\partial t}\Big)_\text{coll}$:
\begin{equation}
\frac{d f}{d t}=\Big(\frac{\partial f}{\partial t}\Big)_\text{coll}.
\end{equation}
In a homogeneous sample with a steady external perturbation, there are no position $\mathbf{r}$ nor time $t$  dependence in the distribution function $f$. Then, $\frac{d f}{d t}$ simplifies to 
\begin{equation}
\frac{d f}{d t}=\dot{\mathbf{k}}\cdot\frac{\partial f_\mathbf{k}}{\partial \mathbf{k}},
\end{equation}
where we have included a subscript $\mathbf{k}$ to the non-equilibrium distribution function $f$ to indicate that it is only a function of $\mathbf{k}$. In a simple relaxation time approximation, the collision integral is replaced by the ratio between the deviation from the equilibrium Fermi-Dirac distribution $f^{(0)}_\mathbf{k}$ and the average time $\tau$ between successive collisions. Hence,
\begin{equation}\label{bte1}
\dot{\mathbf{k}}\cdot\frac{\partial f_\mathbf{k}}{\partial \mathbf{k}}=-\frac{g_\mathbf{k}}{\tau},
\end{equation}
where 
$g_\mathbf{k}=f_\mathbf{k}-f^{(0)}_\mathbf{k}$. Here we assume a constant transport relaxation time $\tau$ in momentum $\mathbf{k}$ and magnetic field $\mathbf{B}$ for a given chemical potential. Note that when we fully consider the collision integral in the system with a nontrivial Berry curvature, the transport relaxation time $\tau$ may show a field dependent anisotropy induced by the coupling between the Berry curvature and magnetic field  \cite{PARK2021}. However, in weak magnetic field regime, we can assume $\tau$ as a constant. Furthermore, when short-range scattering is dominant or charged impurities are fully screened, we can assume that $\tau$ does not depend on $\mathbf{k}$ \cite{XIAO2005, Ashcroft1976, Ziman1960}.

Here we provide a closed-form expression for magnetoconductivity in the presence of both a non-trivial Berry curvature and the Lorentz force within the semiclassical Boltzmann approach. The semiclassical equations of motion for electrons with a charge $q=-e$ in the presence of the Berry curvature are given by \cite{XIAO2010}
\begin{subequations}\label{eom1}
\begin{align}
\dot{\mathbf{r}}&=\frac{\partial \epsilon_\mathbf{m}(\mathbf{k})}{\hbar\partial \mathbf{k}}-\dot{\mathbf{k}}\times\mathbf{\Omega}_\mathbf{k},
\\
\hbar\dot{\mathbf{k}}&=q\mathbf{E}+\frac{q}{c}\dot{\mathbf{r}}\times\mathbf{B},
\end{align}
\end{subequations}
where $\epsilon_\mathbf{m}(\mathbf{k})=\epsilon_{0}(\mathbf{k})-\mathbf{m}_{\mathbf{k}}\cdot\mathbf{B}$, $\epsilon_{0}(\mathbf{k})$ is the unperturbed band energy, $\bf{m}_\mathbf{k}$ is an orbital magnetic moment that couple to the magnetic field  and $\mathbf{\Omega}_\mathbf{k}$ is the Berry curvature. 
It has been reported that disorder affects not only the carrier distribution but also the semiclassical equations of motion, generating a correction to the velocity proportional to the disorder strength \cite{ATENCIA2022}. In this work, we neglect this correction assuming a weak disorder potential for simplicity.
In the presence of a magnetic field, $\dot{\mathbf{r}}$ and $\dot{\mathbf{k}}$ in Eq.~\eqref{eom1} are coupled through the Lorentz force. Combining these two equations of motion, we have
\begin{subequations}\label{eom2}
\begin{align}
\dot{\mathbf{r}}&=D^{-1}_\mathbf{k}[\mathbf{v}_\mathbf{k}-\frac{q}{\hbar}\mathbf{E}\times\mathbf{\Omega}_\mathbf{k}-\frac{q}{\hbar c}({\Omega}_\mathbf{k}\cdot\mathbf{v}_\mathbf{k})\mathbf{B}],
\\
\hbar\dot{\mathbf{k}}&=D^{-1}_\mathbf{k}[q\mathbf{E}+\frac{q}{c}\mathbf{v}_\mathbf{k}\times\mathbf{B}-\frac{q^2}{\hbar c}(\mathbf{E}\cdot\mathbf{B})\mathbf{\Omega}_\mathbf{k}]\label{eqmmomentum},
\end{align}
\end{subequations}
where $\mathbf{v}_\mathbf{k}=\frac{1}{\hbar}\frac{\partial \epsilon_\mathbf{m}(\mathbf{k})}{\partial \mathbf{k}}$ and $D_\mathbf{k}=1-\frac{q}{\hbar c}(\mathbf{\Omega}_\mathbf{k}\cdot\mathbf{B})$ represents the modified density of states in the phase space due to the Berry curvature effect \cite{XIAO2005}. The first term in the square bracket in Eq.~\eqref{eqmmomentum} corresponds to electric force due to an electric field $\mathbf{E}$ and the second term  represents Lorentz force due to a magnetic field $\mathbf{B}$. The last term is anomalous electromagnetic force due to the Berry curvature which leads to positive magnetoconductivity in topological materials.

Replacing $f_\mathbf{k} = f^0_\mathbf{k}+g_\mathbf{k}$ according to the relaxation time approximation in Eq.~\eqref{bte1}, we have
\begin{equation}\label{rta1}
\begin{aligned}
\hbar \dot{\mathbf{k}}\cdot \frac{1}{\hbar}\frac{\partial (f^{(0)}_\mathbf{k}+g_\mathbf{k})}{\partial \mathbf{k}}=-\frac{g_\mathbf{k}}{\tau}.
\end{aligned}
\end{equation}
Plugging Eq.~\eqref{eom2} into Eq.~\eqref{rta1}, we get
\\
\begin{eqnarray}\label{rta2}
D^{-1}_\mathbf{k}\Big[&&q\mathbf{E}\cdot \mathbf{v}_\mathbf{k}\frac{\partial f^{(0)}_\mathbf{k}}{\partial \epsilon_\mathbf{k}}+\frac{q}{\hbar c}\left(\mathbf{v}_\mathbf{k} \times \mathbf{B} \cdot \frac{\partial g_\mathbf{k}}{\partial \mathbf{k}}\right) \nonumber\\ 
&&-\frac{q^2}{\hbar c}(\mathbf{E} \cdot \mathbf{B})(\Omega_\mathbf{k} \cdot \mathbf{v}_\mathbf{k})\frac{\partial f^{(0)}_\mathbf{k}}{\partial \epsilon_\mathbf{k}}\Big]
=-\frac{g_\mathbf{k}}{\tau}.
\end{eqnarray}

In previous studies \cite{DAI2017, KIM2014, NANDY2017} of magnetoconductivity and planar Hall conductivity in topological materials, Lorentz force effect has been often neglected. Therefore, in order to better understand positive longitudinal magnetoconductivity and angle-dependent planar Hall conductivity in topological materials, we take all the terms in Eq.~\eqref{rta2} into consideration.

\begin{figure}[t]
\includegraphics[width=1\linewidth,height=2.4in]{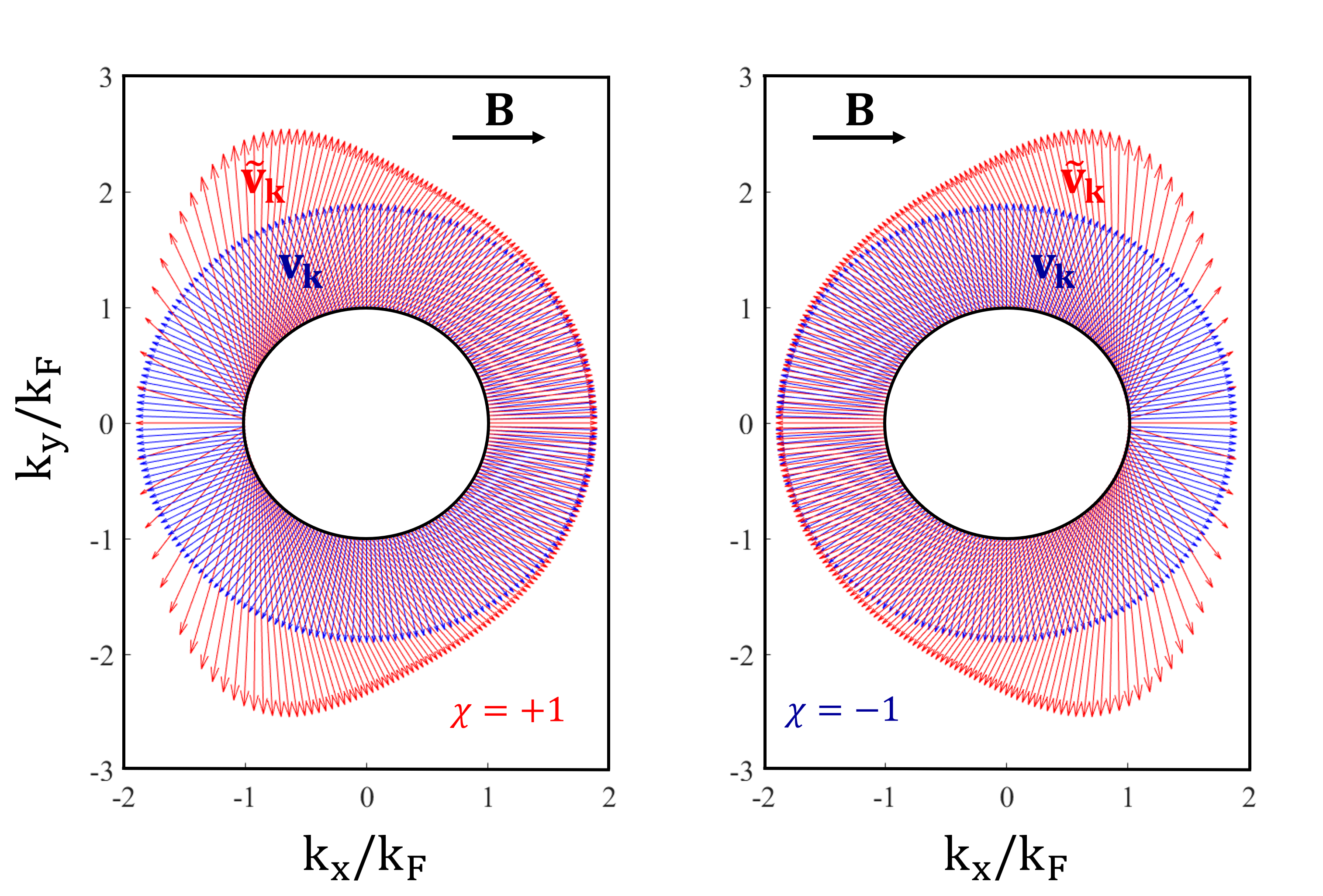}
\caption{Schematic picture comparing $\mathbf{v}_\mathbf{k}$ (blue) and $\tilde{\mathbf{v}}_\mathbf{k}$ (red) near the Weyl nodes described by the Hamiltonian $H=\chi\hbar v_{\rm F}\mathbf{k}\cdot\bm{\sigma}$ for 
$\chi=\pm 1$ in the presence of a constant magnetic field $\mathbf{B}=B_x\hat{x}$. Here the length of an arrow represents the velocity normalized by $v_{\rm F}$.
}  
\label{fig1}
\end{figure}

We can rewrite Eq.~\eqref{rta2} in the following form:
\begin{equation}\label{rta3}
q\mathbf{E}\cdot\tilde{\mathbf{v}}_\mathbf{k}\frac{\partial f^{(0)}_\mathbf{k}}{\partial \epsilon_\mathbf{k}}+\frac{q}{\hbar c}(\tilde{\mathbf{v}}_\mathbf{k}\times\mathbf{B})\cdot\frac{\partial g_\mathbf{k}}{\partial \mathbf{k}}=-\frac{g_\mathbf{k}}{\tau},
\end{equation}
where 
\begin{equation}\label{tildev}
\tilde{\mathbf{v}}_\mathbf{k}={D_\mathbf{k}}^{-1}\Big[\mathbf{v}_\mathbf{k}-\frac{q}{\hbar c}(\mathbf{v}_\mathbf{k}\cdot\mathbf{\Omega}_\mathbf{k})\mathbf{B}\Big].
\end{equation}
Figure \ref{fig1} shows a schematic picture of $\tilde{\mathbf{v}}_\mathbf{k}$ in WSMs in the presence of a magnetic field $\mathbf{B}$. As shown in Fig.~\ref{fig1}, $\tilde{\mathbf{v}}_\mathbf{k}$ is greater in magnitude than $v_{\mathbf{k}}$ in all $\mathbf{k}$ except when $\mathbf{v}_\mathbf{k}$ is parallel to $\mathbf{B}$ giving $\tilde{\mathbf{v}}_\mathbf{k}=\mathbf{v}_\mathbf{k}$.
In general, it is this effect due to Berry curvature that leads to enhanced magnetoconductivity in topological materials.

Our goal now is to solve for $g_\mathbf{k}$ in Eq.~\eqref{rta3}. We assume that a small deviation from the equilibrium distribution takes the form of $g_\mathbf{k}=\tilde{\mathbf{v}}_\mathbf{k}\cdot\mathbf{G}$ where $\mathbf{G}$ is some arbitrary vector that is independent of $\tilde{\mathbf{v}}_\mathbf{k}$. Upon plugging $g_\mathbf{k}=\tilde{\mathbf{v}}_\mathbf{k}\cdot\mathbf{G}$ into Eq.~\eqref{rta3}, we obtain 
\begin{equation}
q\tau\frac{\partial f^{(0)}_\mathbf{k}}{\partial \epsilon_\mathbf{k}}\mathbf{E}\cdot\tilde{\mathbf{v}}_\mathbf{k}+\frac{q\tau}{c}(\tilde{\mathbf{v}}_\mathbf{k}\times\mathbf{B})\cdot\mathds{M}^{-1}\mathbf{G}+\tilde{\mathbf{v}}_\mathbf{k}\cdot\mathbf{G}=0,
\end{equation}
where $\mathds{M}^{-1}_{ij}=\frac{1}{\hbar}\frac{\partial \tilde{\rm v}_{\mathbf{k},j}}{\partial k_i}$ is the inverse mass tensor. Factoring $\tilde{\mathbf{v}}_\mathbf{k}$ out,
\begin{equation}\label{eqgk}
\begin{aligned}
g_\mathbf{k}=&\tilde{\mathbf{v}}_\mathbf{k}\cdot\bigg[q\tau\Big(-\frac{\partial f^{(0)}_\mathbf{k}}{\partial \epsilon_\mathbf{k}}\Big)\mathbf{E}-\frac{q\tau}{c}(\mathbf{B}\times\mathds{M}^{-1}\mathbf{G})\bigg].
\end{aligned}
\end{equation}
From $g_\mathbf{k}=\tilde{\mathbf{v}}_\mathbf{k}\cdot\mathbf{G}$ and Eq.~\eqref{eqgk}, we obtain a self-consistent form of $\mathbf{G}$ as
\begin{equation}\label{eqG}
\mathbf{G}=\mathbf{G}_0-\frac{q\tau}{c}(\mathbf{B}\times\mathds{M}^{-1}\mathbf{G}),
\end{equation}
where 
\begin{equation}
\mathbf{G}_0=q\tau\Big(-\frac{\partial f^{(0)}_\mathbf{k}}{\partial \epsilon_\mathbf{k}}\Big)\mathbf{E}.
\end{equation}
The solution for Eq.~\eqref{eqG} can be obtained as
\begin{equation}\label{defG}
\mathbf{G}=q\tau\Big(-\frac{\partial f^{(0)}_\mathbf{k}}{\partial \epsilon_\mathbf{k}}\Big)\Big(\mathds{1}+\frac{q\tau}{c}\mathds{F}\mathds{M}^{-1}\Big)^{-1}\mathbf{E},
\end{equation}
where $\mathds{F}_{ij}=\sum_i \epsilon_{ijk} B_k$ is the magnetic field strength tensor (see Appendix \ref{A} for a detailed derivation). Thus, we obtain
\begin{equation}\label{eq15}
g_\mathbf{k}=q\tau\Big(-\frac{\partial f^{(0)}_\mathbf{k}}{\partial \epsilon_\mathbf{k}}\Big)\tilde{\mathbf{v}}_\mathbf{k}\cdot\Big[\Big(\mathds{1}+\frac{q\tau}{c}\mathds{F}\mathds{M}^{-1}\Big)^{-1}\mathbf{E}\Big].
\end{equation}
The current is therefore computed as
\begin{equation}
\begin{aligned}
J_\alpha=q\int\frac{d^d k}{(2\pi)^d}D_\mathbf{k}f_{\mathbf{k}}v_\alpha,
\end{aligned}
\end{equation}
where $\alpha$ is the direction in which we measure the current. Let us denote the current associated with $f^{(0)}_\mathbf{k}$ as $J^\text{int}_\alpha=q\int\frac{d^d k}{(2\pi)^d}D_\mathbf{k}f^{(0)}_{\mathbf{k}} {{v}}_\alpha$ and that associated with $g_\mathbf{k}$ as $J^\text{ext}_\alpha=q\int\frac{d^d k}{(2\pi)^d}D_\mathbf{k}g_\mathbf{k}{{v}}_\alpha$. From now on, we will only focus on the extrinsic contribution $J^\text{ext}_\alpha$ arising from scatterings.

Plugging Eq.~\eqref{eq15} into $J^\text{ext}_\alpha$ and using $J_{\alpha}^{\text{ext}}=\sum_{\beta} \sigma_{\alpha\beta} E_\beta$ relation, we finally arrive at
\begin{eqnarray}\label{eq17}
\sigma_{\alpha\beta}=q^2\int&&\frac{d^dk}{(2\pi)^d} D_\mathbf{k}\tau\Big(-\frac{\partial f^{(0)}_\mathbf{k}}{\partial \epsilon_\mathbf{k}}\Big)\tilde{v}_\alpha \nonumber\\
&&\times\Big(\tilde{\mathbf{v}}_\mathbf{k}\cdot\Big\{\Big[\mathds{1}+\frac{q\tau}{c}\mathds{F}\mathds{M}^{-1}\Big]^{-1}\hat{\beta}\Big\}\Big),
\end{eqnarray}
where $\beta$ is the direction of an electric field. 

Now let us assume that the mobility tensor $e\tau\mathds{M}^{-1}$ is set to a constant $\mu$ for simplicity. Then $\mathbf{G}$ in Eq.~\eqref{defG} becomes (see Appendix \ref{A})
\begin{equation}
\mathbf{G}=q\tau\Big(-\frac{\partial f^{(0)}_\mathbf{k}}{\partial \epsilon_\mathbf{k}}\Big)\frac{\mathbf{E}+\frac{\mu}{c}\mathbf{E}\times\mathbf{B}+\frac{\mu^2}{c^2}(\mathbf{E}\cdot\mathbf{B})\mathbf{B}}{1+\frac{\mu^2}{c^2}|\mathbf{B}|^2}.
\end{equation}
Note that the obtained $\mathbf{G}$ is consistent with the assumption that $\mathbf{G}$ is independent of $\tilde{\mathbf{v}}_\mathbf{k}$. Then $g_\mathbf{k}$ is given by
\begin{equation}\label{eq11}
g_\mathbf{k}=q\tau\Big(-\frac{\partial f^{(0)}_\mathbf{k}}{\partial \epsilon_\mathbf{k}}\Big)\tilde{\mathbf{v}}_\mathbf{k}\cdot\frac{\mathbf{E}+\frac{\mu}{c}\mathbf{E}\times\mathbf{B}+\frac{\mu^2}{c^2}(\mathbf{E}\cdot\mathbf{B})\mathbf{B}}{1+\frac{\mu^2}{c^2}|\mathbf{B}|^2}.
\end{equation}

Using Eq.~\eqref{eq11}, we finally obtain the following form for magnetoconductivity
\begin{eqnarray}
\label{eqmucon}
\sigma_{\alpha\beta}=&&q^2\int\frac{d^dk}{(2\pi)^d}\frac{D_\mathbf{k}\tau(-\frac{\partial f^{(0)}_\mathbf{k}}{\partial \epsilon_\mathbf{k}})}{1+\frac{\mu^2}{c^2}|\mathbf{B}|^2}\\
&&\times\Big[\tilde{v}_\alpha\tilde{v}_\beta-\frac{\mu}{c}\tilde{v}_\alpha(\tilde{\mathbf{v}}_\mathbf{k}\times\mathbf{B})_\beta+\frac{\mu^2}{c^2}(\tilde{\mathbf{v}}_\mathbf{k}\cdot\mathbf{B})\tilde{v}_\alpha B_\beta\Big].\nonumber
\end{eqnarray}
The above form is a general expression of magnetoconductivity for topological materials including the Berry curvature and Lorentz force within the semiclassical regime under the assumption that $\mathbf{G}$ in $g_\mathbf{k}=\tilde{\mathbf{v}}_\mathbf{k}\cdot\mathbf{G}$ is independent of $\tilde{\mathbf{v}}_\mathbf{k}$ and the mobility tensor is a constant.


\section{Magnetotransport in Weyl semimetals}

In this section, we study the magnetotransport properties of WSMs in three dimensions using a closed-form expression for magnetoconductivity Eq.~\eqref{eqmucon} discussed in the previous section. For simplicity, we consider a single  Weyl node described by the Hamiltonian $H=\chi\hbar v_{\rm F}\mathbf{k}\cdot\bm{\sigma}$ which has isotropic linear dispersion, where 
$\chi=\pm 1$ are for the different chiralities of Weyl fermions and 
$\bm{\sigma}$ are the Pauli matrices.

\subsection{Longitudinal magnetoconductivity}

To investigate longitudinal magnetoconductivity $\sigma_{xx}(\mathbf{B})$  in WSMs, without loss of generality, we set the electric and magnetic field orientations as $\mathbf{E}=E_x\hat{x}$ and $\mathbf{B}=B_x\hat{x}+B_y\hat{y}$, respectively. Organizing terms in Eq.~\eqref{eqmucon} in powers of $\mu$ and using Eq.~\eqref{tildev},
\begin{equation}\label{eqsimplesum}
\sigma_{xx}=q^2\int\frac{d^3k}{(2\pi)^3}\frac{\tau(-\frac{\partial f^{(0)}_\mathbf{k}}{\partial \epsilon_\mathbf{k}})D^{-1}_\mathbf{k}}{1+\frac{\mu^2}{c^2}|\mathbf{B}|^2}\Big[\Sigma_{\mathbf k}^{(0)}+\Sigma_{\mathbf k}^{(1)}+\Sigma_{\mathbf k}^{(2)}\Big],
\end{equation}
where $\Sigma_{\mathbf k}^{(i)}$ is a sum of the terms that include $i$th order of $\mu$. Due to the $D^{-1}_{\mathbf{k}}$ term in Eq.~\eqref{eqsimplesum}, it is difficult to obtain an analytic expression for $\sigma_{xx}$ incorporating the full density of states correction. Therefore, to obtain a simple closed form result, we first assume  $D^{-1}_{\mathbf{k}}$ as $1$ in Eq.~\eqref{eqsimplesum}.  We will discuss the correction beyond this approximation later. Here, $\Sigma_{\mathbf k}^{(i)}$ are defined as
\begin{subequations}\label{eqsimplesum-Sigma}
\begin{alignat}{4}
\Sigma_{\mathbf k}^{(0)}=&~v^2_x-2\frac{q}{\hbar c}v_x(\mathbf{v}_{\mathbf{k}}\cdot\mathbf{\Omega}_\mathbf{k})|\mathbf{B}|\cos\Gamma\label{Sigma1}\nonumber\\&+\frac{q^2}{(\hbar c)^2}(\mathbf{v}_{\mathbf{k}}\cdot\mathbf{\Omega}_\mathbf{k})^2|\mathbf{B}|^2\cos^2\Gamma,\\
\nonumber\\
\Sigma_{\mathbf k}^{(1)}=&~\frac{\mu}{c}\Big[-v_x(\mathbf{v}_\mathbf{k}\times\mathbf{B})_x\label{Sigma2}\nonumber\\&~+\frac{q}{\hbar c}(\mathbf{v}_\mathbf{k}\times\mathbf{B})_x(\mathbf{v}_\mathbf{k}\cdot\mathbf{\Omega}_\mathbf{k})|\mathbf{B}|\cos\Gamma\Big],\\
\nonumber\\
\Sigma_{\mathbf k}^{(2)}=&~
\frac{\mu^2}{c^2}\Big[(\mathbf{v}_\mathbf{k}\cdot\mathbf{B})v_x|\mathbf{B}|\cos\Gamma\nonumber\\&~-\frac{q}{\hbar c}(\mathbf{v}_\mathbf{k}\cdot\mathbf{\Omega}_\mathbf{k})|\mathbf{B}|^3v_x\cos\Gamma\nonumber\\&~
-\frac{q}{\hbar c}(\mathbf{v}_\mathbf{k}\cdot\mathbf{B})(\mathbf{v}_\mathbf{k}\cdot\mathbf{\Omega}_\mathbf{k})|\mathbf{B}|^2\cos^2\Gamma\nonumber\\&~+\frac{q^2}{(\hbar c)^2}(\mathbf{v}_\mathbf{k}\cdot\mathbf{\Omega}_\mathbf{k})^2|\mathbf{B}|^4\cos^2\Gamma\Big],
\end{alignat}
\end{subequations}
where $\Gamma$ is the angle between $\bf E$ and $\bf B$. The first term $v^2_x$ in Eq.~\eqref{Sigma1} gives a well known longitudinal conductivity in the absence of magnetic field: $\sigma_{xx}(\mathbf{B}=0)=q^2 N_0D\equiv\sigma_0$ where ${N_0}$ is the density of states at the Fermi energy and $D=v_{\rm F}^2 \tau/d$ is the diffusion constant with $d=3$.

Collecting terms that would give us non-zero contribution after momentum integral, Eq.~\eqref{eqsimplesum} can be rewritten as
\begin{eqnarray}\label{eq101}
&&\sigma_{xx}=q^2\int\frac{d^3k}{(2\pi)^3}\frac{\tau(-\frac{\partial f^{(0)}}{\partial \epsilon_\mathbf{k}})}{(1+\frac{\mu^2}{c^2}|\mathbf{B}|^2)}\\
&&\times\bigg[v^2_x+\frac{q^2}{(\hbar c)^2}(\mathbf{v}_\mathbf{k}\cdot\mathbf{\Omega}_\mathbf{k})^2|\mathbf{B}|^2\cos^2\Gamma\nonumber\\
&&+\frac{\mu^2}{c^2}v^2_x|\mathbf{B}|^2\cos^2\Gamma+\frac{\mu^2}{c^2}\frac{q^2}{(\hbar c)^2}(\mathbf{v}_\mathbf{k}\cdot\mathbf{\Omega}_\mathbf{k})^2|\mathbf{B}|^4\cos^2\Gamma\bigg].\nonumber
\end{eqnarray}

The Berry curvature in an isotropic WSM is $\mathbf{\Omega}_\mathbf{k}=\chi\frac{\mathbf{k}}{2 |\mathbf{k}|^3}$. Therefore, $\mathbf{v}_\mathbf{k}\cdot\mathbf{\Omega}_\mathbf{k}$ becomes $\chi{v_{\rm F}}/({2 k_{\rm F}^2})$ where $v_{\rm F}$ and $k_{\rm F}$ are the Fermi velocity and Fermi wave vector, respectively. Note that all the surviving terms are even functions of magnetic field $|\mathbf{B}|$ and independent of $\chi$. Here we emphasize that there are two kinds of magnetic field effects: the Lorentz force and anomalous velocity effect due to the Berry curvature. The terms that are related to the Lorentz force comes with a $\frac{\mu}{c}|\mathbf{B}|$ factor. On the other hand, the terms that are related to the Berry curvature comes with a $\frac{q}{\hbar c}|\mathbf{\Omega}_\mathbf{k}||\mathbf{B}|=\frac{q^2}{(\hbar c)^2}\frac{1}{2k_{\rm F}^2}$ factor.

Since the terms coupled with the magnetic field are proportional to either $\frac{\mu}{c}|\mathbf{B}|$ or $\frac{q}{\hbar c}|\mathbf{\Omega}_\mathbf{k}||\mathbf{B}|$, we introduce the following dimensionless parameters:
\begin{subequations}
\begin{eqnarray}
b_{\mu}&\equiv&\frac{\mu}{c}|\mathbf{B}|\equiv\frac{q\tau}{\hbar c} \frac{v_{\rm F}}{k_{\rm F}}|\mathbf{B}|, \\ b_{\rm BC}&\equiv&~\frac{q}{\hbar c}|\mathbf{\Omega}_\mathbf{k}||\mathbf{B}|~\equiv\frac{q}{\hbar c}\frac{1}{2 k_{\rm F}^2}|\mathbf{B}|.
\end{eqnarray}
\end{subequations}
Note that these two dimensionless parameters are related with each other as $b_{\mu}/b_{\rm BC}=2k_{\rm F}l$, where $l=v_{\rm F} \tau$ is the mean-free path. This gives us an important insight that the Lorentz force effect cannot be simply neglected when studying magnetotransport in topological materials within the semiclassical Boltzmann approach which is valid in  $k_{\rm F} l\gg1$ regime.

\begin{figure}[t]
\includegraphics[width=1\linewidth,height=2.8in]{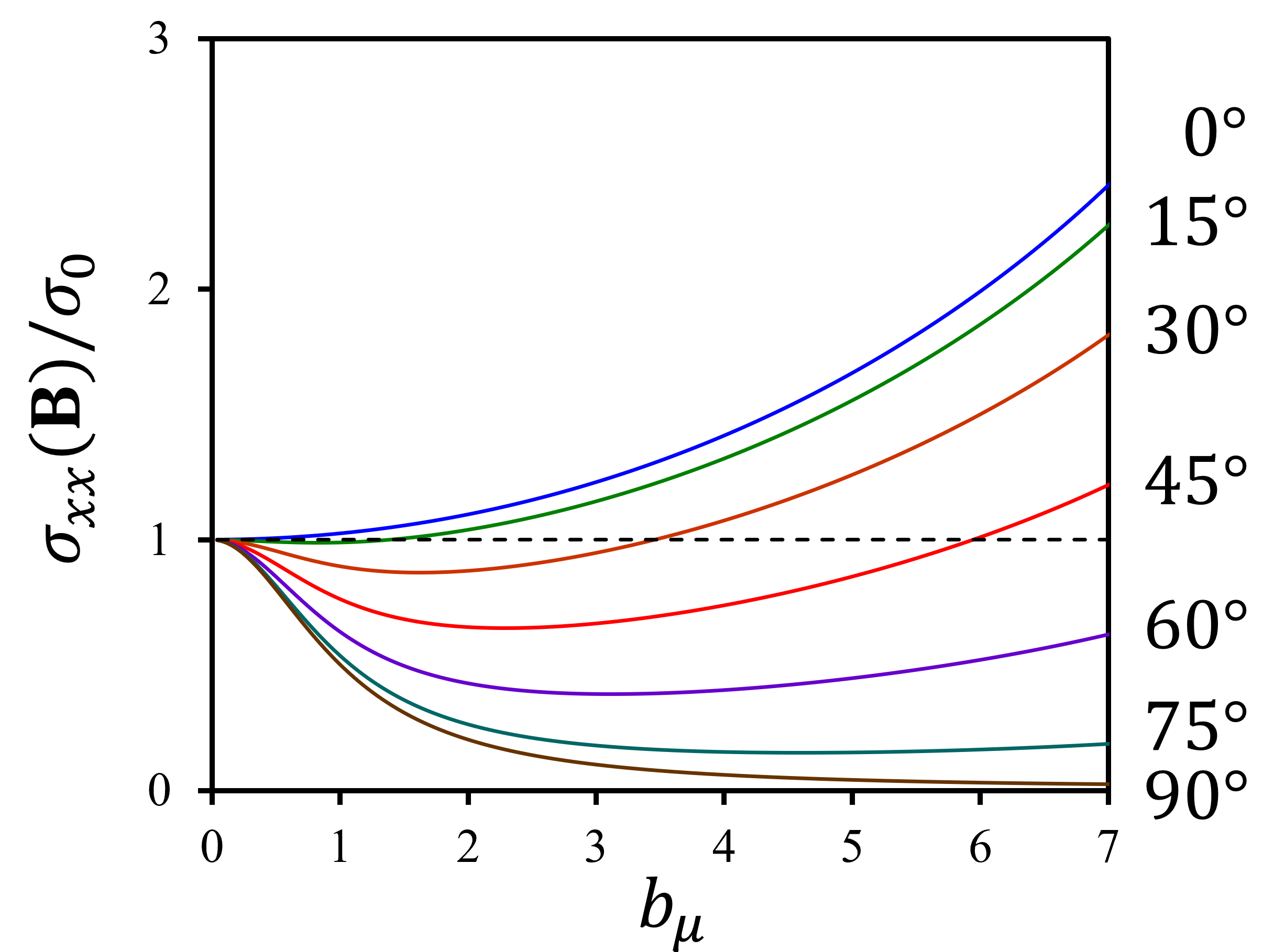}
\caption{Longitudinal conductivity as a function of $b_{\mu}$ at selected angles between $\mathbf{E}$ and $\mathbf{B}$ in WSMs for ${k}_{\rm F}l=4$.}
\label{fig2}
\end{figure}

Carrying out the momentum integral in Eq.~\eqref{eq101}, we obtain a longitudinal magnetoconductivity in WSMs for an arbitrary external magnetic field under the assumption that $D^{-1}_{\mathbf{k}}=1$ (see Appendix \ref{B1}):
\begin{equation}\label{eqAformsigmaxxDk1}
\sigma_{xx}(\mathbf{B})=\sigma_0\Big[\frac{1}{1+b^2_{\mu}}+\Big({\frac{b^2_{\mu}}{1+b^2_{\mu}} +3 b^2_{\rm BC}}\Big)\cos^2{\Gamma}\Big].
\end{equation}
Note that $\sigma_{xx}(\mathbf{B})$ in Eq.~\eqref{eqAformsigmaxxDk1} is independent of $\chi$, as $b_\text{BC}$ comes with a power of two. Thus, the contribution from multiple degenerate Weyl nodes enters as a degeneracy factor when internode scatterings are neglected. 
Most of previous works for longitudinal magnetoconductivity in WSMs using Boltzmann transport theory reported $\sigma_{xx}(\Gamma)\sim\sigma_0+\Delta\sigma\cos^2{\Gamma}$  where $\Delta\sigma$ is the additional positive magnetoconductivity that scales quadratically in magnetic field \cite{KIM2014, DAI2017, NANDY2017, IMRAN2018}. Instead, Eq.~\eqref{eqAformsigmaxxDk1} shows a non-monotonic behavior of magnetoconductivity in magnetic field due to the competition between the Berry curvature and Lorentz force depending on $k_{\rm F} l$, as shown in Fig.~\ref{fig2}. This result is consistent with the previous calculation for WSMs obtained by expanding the non-equilibrium distribution function in Fourier harmonics \cite{IMRAN2018}.

We now come back to the approximation we made: $D^{-1}_{\mathbf{k}}=1$. Note that for weak magnetic fields, we could Taylor expand $D^{-1}_{\mathbf{k}}$ as
\begin{equation}\label{Taylor}
D^{-1}_\mathbf{k}=1+\frac{q}{\hbar c}(\mathbf{B}\cdot\mathbf{\Omega}_\mathbf{k})+(\frac{q}{\hbar c})^2(\mathbf{B}\cdot\mathbf{\Omega}_\mathbf{k})^2+\cdots.
\end{equation}

We emphasize that earlier studies of magnetotransport also took the $D_\mathbf{k}$ correction into account but incompletely. For instance, Kim {\textit {et al}.}~\cite{KIM2014} only took even terms in Taylor expanded $D^{-1}_\mathbf{k}$ in Eq.~\eqref{Taylor} which gave rise to an additional positive correction to magnetoconductivity described in Eq.~\eqref{eqAformsigmaxxDk1}. However, numerical calculations of the full longitudinal magnetoconductivity [Eq.~\eqref{eqsimplesum}] show a reduced value compared to Eq.~\eqref{eqAformsigmaxxDk1}. The reduced magnetoconductivity is due to terms that are odd orders in $|\mathbf{B}|$ in Eq.~\eqref{eqsimplesum} coupling with odd orders in $|\mathbf{B}|$ terms in the Taylor expanded $D^{-1}_\mathbf{k}$ in Eq.~\eqref{Taylor}. As a result, the pairs of odd terms in $|\mathbf{B}|$ give rise to non-vanishing even terms in $|\mathbf{B}|$ which additionally give negative corrections to the magnetoconductivity.   

By incorporating first three terms in Eq.~\eqref{Taylor}, we obtain the longitudinal magnetoconductivity in WSMs up to quadratic order in $b_{\rm BC}^2$ as
\begin{eqnarray}
\sigma_{xx}(\mathbf{B})&&\approx\frac{\sigma_0}{1+b_{\mu}^2}\\&&\times \Big\{1+\frac{1}{5}b^2_{\rm BC}+\Big[\frac{7}{5}b^2_{\rm BC}+b_{\mu}^2\Big(1+\frac{8}{5}b^2_{\rm BC}\Big)\Big]\cos^2{\Gamma}\Big\}.\nonumber
\end{eqnarray}
See Appendix \ref{B1} for detailed derivations.
This result is well matched with the previous work which focused on specific angles between applied electric and magnetic fields \cite{IMRAN2018}.

\subsection{Planar Hall conductivity}

\begin{figure}[t]
\includegraphics[width=1\linewidth]{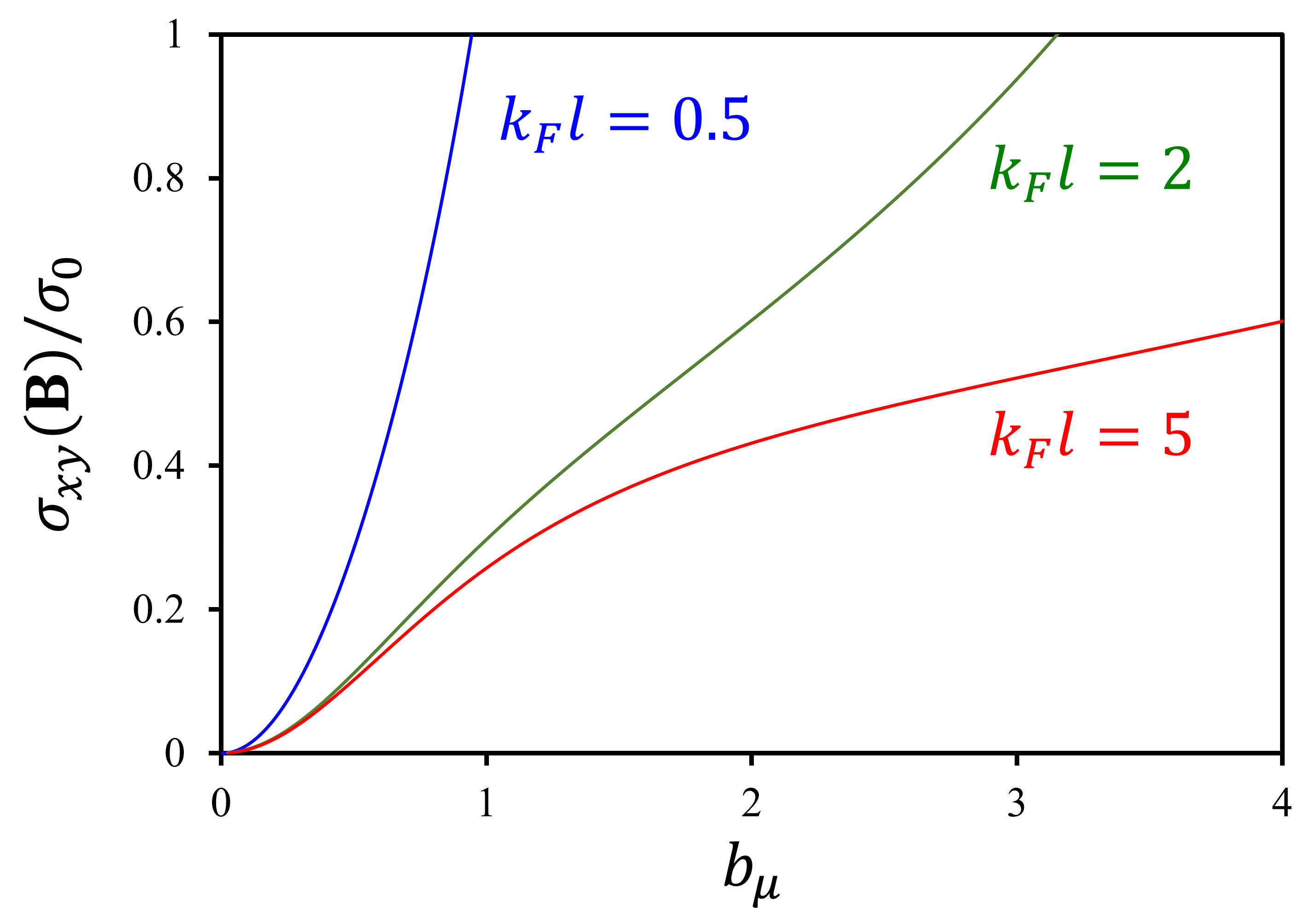}
\caption{Planar Hall conductivity as a function of $b_{\mu}$ in WSMs at $\Gamma=\frac{\pi}{4}$ for $k_{\rm F}l=0.5, 2, 5$.}
\label{fig3}
\end{figure}

To investigate the planar Hall conductivity in WSMs, we again set the electric and magnetic field orientations as $\mathbf{E}=E_x\hat{x}$ and $\mathbf{B}=B_x\hat{x}+B_y\hat{y}$, respectively. Neglecting terms that would give zero contribution to conductivity, Eq.~\eqref{eqmucon} gives the following form of the planar Hall conductivity $\sigma_{yx}(\mathbf{B})$ assuming $D^{-1}_{\mathbf{k}}=1$ (see Appendix \ref{B2}):
\begin{equation}\label{eq300}
\sigma_{yx}(\mathbf{B})=\sigma_0\Big[\Big( {\frac{b^2_{\mu}}{1+b^2_{\mu}} +3 b^2_{\rm BC}} \Big)\sin{\Gamma}\cos{\Gamma}\Big].
\end{equation}
Note that $\sigma_{yx}(\mathbf{B})$ in Eq.~\eqref{eq300} is independent of $\chi$. Equation \eqref{eq300} is the analytic form of the planar Hall conductivity in WSMs for an arbitrary angle of the applied magnetic field $|\mathbf{B}|$. The angle dependence of the planar Hall conductivity is well matched with the previous works $\sigma_{yx}(\Gamma)\sim\sin{\Gamma}\cos{\Gamma}$ \cite{BURKOV2017, NANDY2017, LI2018, LIANG2019, LIU2019, MA2019, LI2020} but the field dependence shows an additional contribution from the Lorentz force in addition to a quadratic field dependence due to the Berry curvature. As shown in Fig~\ref{fig3}, the planar Hall conductivity shows different $|\mathbf{B}|$ dependence at different $k_{\rm F} l$ regimes. For large $k_{\rm F} l$, the planar Hall conductivity roughly increases with $|\mathbf{B}|$ quadratically at low $|\mathbf{B}|$ field and saturates at high $|\mathbf{B}|$ field. For low $k_{\rm F} l$ regime, the planar Hall conductivity shows $|\mathbf{B}|^2$ dependence with no sign of saturation in a broad range of magnetic field as reported in the previous studies \cite{NANDY2017}.

We now come back to the approximation we made: $D^{-1}_{\mathbf{k}}=1$. Again, including the Taylor expanded $D^{-1}_{\mathbf{k}}$ in Eq.~\eqref{Taylor}, we obtain the planar Hall magnetoconductivity in WSMs as 
\begin{equation}
\sigma_{yx}(\mathbf{B})\approx\sigma_0\Big[\frac{b^2_{\mu}}{1+b^2_{\mu}}\Big(1+\frac{1}{5}b^{2}_{\rm BC}\Big)+\frac{7}{5}b^{2}_{\rm BC}\Big]\sin{\Gamma}\cos{\Gamma}.
\end{equation}
See Appendix \ref{B2} for detailed derivations.
\\
\\
\subsection{Hall conductivity}

To investigate the Hall conductivity in WSMs, we set the electric and magnetic field orientations as $\mathbf{E}=E_x\hat{x}$ and $\mathbf{B}=B_y\hat{y}$, respectively. Then  Eq.~\eqref{eqmucon} gives the following form of Hall conductivity $\sigma_{zx}(\mathbf{B})$ assuming $D^{-1}_{\mathbf{k}}=1$ (see Appendix B 3):

\begin{equation}\label{eq301}
\sigma_{zx}(\mathbf{B})=\sigma_0\Big(\frac{b_{\mu}}{1+b^2_{\mu}}\Big).
\end{equation}

Note that this result is identical to the conventional magnetotransport result. This implies that there is no anomalous velocity effect in the Hall conductivity.

We now come back to the approximation we made: $D^{-1}_{\mathbf{k}}=1$. Again, including the Taylor expanded $D^{-1}_{\mathbf{k}}$ in Eq.~\eqref{Taylor}, we can obtain the Hall magnetoconductivity in WSMs as 
\begin{equation}
\sigma_{zx}(\mathbf{B})\approx\sigma_0\frac{b_\mu}{1+b^2_\mu}\left(1+\frac{1}{5}b^2_\text{BC}\right).
\end{equation}
See Appendix \ref{B3} for detailed derivations.

\subsection{Conductivity and resistivity tensors}

Combining the previous results of magnetoconductivity in WSMs, we obtain the following conductivity tensor under $\mathbf{B}=B_x\hat{x}+B_y\hat{y}$ with $D^{-1}_{\mathbf{k}}=1$:
\begin{widetext}
\begin{equation}\begin{aligned}\label{conductivitytensor1}
{\bm{\sigma}(\mathbf{B})}=\sigma_0 \begin{bmatrix} \frac{1}{1+b^2_{\mu}}+( {\frac{b^2_{\mu}}{1+b^2_{\mu}} +3 b^2_{\rm BC}} )\cos^2{\Gamma} & ( {\frac{b^2_{\mu}}{1+b^2_{\mu}} +3 b^2_{\rm BC}} )\sin{\Gamma}\cos{\Gamma} & -\frac{b_{\mu}}{1+b^2_{\mu}}\sin{\Gamma} \\ ( {\frac{b^2_{\mu}}{1+b^2_{\mu}} +3 b^2_{\rm BC}} )\sin{\Gamma}\cos{\Gamma} & \frac{1}{1+b^2_{\mu}}+( {\frac{b^2_{\mu}}{1+b^2_{\mu}} +3 b^2_{\rm BC}} )\sin^2{\Gamma} & \frac{b_{\mu}}{1+b^2_{\mu}}\cos{\Gamma} \\ \frac{b_{\mu}}{1+b^2_{\mu}}\sin{\Gamma} & -\frac{b_{\mu}}{1+b^2_{\mu}}\cos{\Gamma} &  \frac{1}{1+b^2_{\mu}} \\ \end{bmatrix},
\end{aligned}\end{equation}
\end{widetext}
where $\Gamma$ is the angle between applied electric and magnetic fields. Since the resistivity tensor $\bm{\rho}(\mathbf{B})$ is the inverse of the conductivity tensor $\bm{\sigma}(\mathbf{B})$, we obtain $\bm{\rho}(\mathbf{B})$ in the following form:
\begin{equation}\label{resistivitytensor1}\begin{aligned}
\bm{\rho}(\mathbf{B})=\rho_0 &\begin{bmatrix} \frac{1+3 b^2_{\rm BC}\sin^2{\Gamma}}{1+3 b^2_{\rm BC}} & -\frac{3 b^2_{\rm BC}\sin{\Gamma}\cos{\Gamma}}{1+3 b^2_{\rm BC}}  & b_{\mu}\sin{\Gamma} \\ -\frac{3 b^2_{\rm BC}\sin{\Gamma}\cos{\Gamma}}{1+3 b^2_{\rm BC}} & \frac{1+3 b^2_{\rm BC}\cos^2{\Gamma}}{1+3 b^2_{\rm BC}} & -b_{\mu}\cos{\Gamma} \\ -b_{\mu}\sin{\Gamma} & b_{\mu}\cos{\Gamma} & 1 \\ \end{bmatrix},
\end{aligned}\end{equation}
\\
where $\rho_0=\rho_{xx}(\mathbf{B}=0)=\sigma_0^{-1}$. Note that although both the Lorentz force and the Berry curvature induced terms are present in $\sigma_{ij}(\mathbf{B})$, the Lorentz force induced terms do not appear in the resistivity $\rho_{ij}(\mathbf{B})$ where $i, j = x, y$. This result agrees with the previous one reporting that $\rho_{xx}=\rho_{\parallel}-\Delta\rho\cos^2{\Gamma}$, $\rho_{xy}=-\Delta\rho\sin{\Gamma}\cos{\Gamma}$ where $\rho_{\parallel}$ ($\rho_{\perp}$) is the resistivity in longitudinal (transverse) magnetic field
and $\Delta\rho=\rho_{\perp}-\rho_{\parallel}=3 b^2_{\rm BC}/(1+3 b^2_{\rm BC})$ is the resistivity anisotropy \cite{,JAN1957, BURKOV2017}. For the conductivity and resistivity tensors including the Taylor expanded $D^{-1}_{\mathbf{k}}$, see Appendix \ref{B4}.



\section{Conclusion}

In this work, we presented a closed-form expression for the magnetoconductivity using the semiclassical magnetotransport theory that fully incorporates the Berry curvature and the Lorentz force effects. We then applied this formula to WSMs and obtained analytic expressions for the longitudinal, planar Hall and Hall conductivities in terms of dimensionless parameters $b_{\mu}$ and $b_{\rm BC}$ which are normalized magnetic fields associated with the Lorentz force and the Berry curvature, respectively. From these results, we showed a non-monotonic field dependence in the longitudinal and planar Hall conductivities depending on $k_{\rm F}l$. Furthermore, we clearly demonstrated that although the Lorentz force effect is manifested in the planar Hall conductivity, its contribution vanishes in the planar Hall resistivity. 


\acknowledgments
This work was supported by the National Research Foundation of Korea (NRF) grant funded by the Korea government (MSIT) (Grant No. 2018R1A2B6007837) and Creative-Pioneering Researchers Program through Seoul National University (SNU).




\appendix
\widetext
\section{Derivation of the nonequilibrium distribution function}\label{A}
Here we go through a detailed derivation of the nonequilibrium distribution function. Let us start with Eq.~\eqref{eqG} in the main text:
\begin{equation}\label{AP1}
\begin{aligned}
\mathbf{G}&=\mathbf{G}_0-\frac{q\tau}{c}(\mathbf{B}\times\mathds{M}^{-1}\mathbf{G}),
\end{aligned}
\end{equation}
where $\mathds{M}^{-1}_{ij}=\frac{1}{\hbar}\frac{\partial \tilde{\rm v}_{\mathbf{k},j}}{\partial k_i}$ is the inverse mass tensor and 
\begin{equation}
\mathbf{G}_0=q\tau\Big(-\frac{\partial f^{(0)}_\mathbf{k}}{\partial \epsilon_\mathbf{k}}\Big)\mathbf{E}.
\end{equation}

To solve $\mathbf{G}$ in Eq.~\eqref{AP1}, note that Eq.~\eqref{AP1} is given by the following self-consistent form for a vector $\mathbf{x}$:
\begin{equation}
\label{eq:self_consistent_form_Mx}
\mathbf{x}=\mathbf{a}+(M\mathbf{x})\times\mathbf{b},
\end{equation}
where $\mathbf{a}$ and $\mathbf{b}$ are vectors and $M$ is a matrix. Then Eq.~\eqref{eq:self_consistent_form_Mx} can be rewritten as
\begin{eqnarray}
x_i=a_i+\sum_{jk} \epsilon_{ijk} (M\mathbf{x})_j b_k 
=a_i+\sum_{jk} F_{ij} M_{jk} x_k,
\end{eqnarray}
where $F_{ij}=\sum_k \epsilon_{ijk} b_k$. Thus, we have
\begin{equation}
\label{eq:self_consistent_form_Mx-solution}
\mathbf{x}=\mathbf{a}+F M \mathbf{x}=(1-FM)^{-1}\mathbf{a}.
\end{equation}

Using the above result Eq.~\eqref{eq:self_consistent_form_Mx-solution}, we obtain $\mathbf{G}$ as the following form:
\begin{equation}\label{AP7}
\mathbf{G}=q\tau\Big(-\frac{\partial f^{(0)}_\mathbf{k}}{\partial \epsilon_\mathbf{k}}\Big)\Big(\mathds{1}+\frac{q\tau}{c}\mathds{F}\mathds{M}^{-1}\Big)^{-1}\mathbf{E},
\end{equation}
where $\mathds{F}_{ij}=\sum_i \epsilon_{ijk} B_k$ is the magnetic field strength tensor. Finally, the nonequilibrium distribution function $g_\mathbf{k}=\tilde{\mathbf{v}}_{\mathbf{k}}\cdot\mathbf{G}$ is given by
\begin{equation}
g_\mathbf{k}=q\tau\Big(-\frac{\partial f^{(0)}_\mathbf{k}}{\partial \epsilon_\mathbf{k}}\Big)\tilde{\mathbf{v}}_\mathbf{k}\cdot\Big[\Big(\mathds{1}+\frac{q\tau}{c}\mathds{F}\mathds{M}^{-1}\Big)^{-1}\mathbf{E}\Big].
\end{equation}

The mobility tensor is defined as
\begin{equation}
\tilde{\mu}_{ij}=e\tau\mathds{M}^{-1}_{ij},
\end{equation}
which is in general a non-diagonal matrix. For a system with an isotropic energy dispersion in the absence of a magnetic field, the mobility tensor is given by a scalar multiple of an identity matrix. For simplicity, we assume that the mobility tensor is set to a constant $\mu$. 
Then, we can rewrite Eq.~\eqref{AP1} as
\begin{equation}\label{AP10}
\begin{aligned}
\mathbf{G}&=\mathbf{G}_0-\frac{\mu}{c}\mathbf{B}\times\mathbf{G}.
\end{aligned}
\end{equation}

Note that Eq.~\eqref{AP10} is given by the following self-consistent form for a vector $\mathbf{x}$:
\begin{equation}
\label{eq:self_consistent_form_x}
\mathbf{x}=\mathbf{a}+\mathbf{x}\times\mathbf{b},
\end{equation}
where $\mathbf{a}$ and $\mathbf{b}$ are vectors. Then Eq.~\eqref{eq:self_consistent_form_x} can be rewritten as
\begin{eqnarray}
x_i&=&a_i+\sum_{jk} \epsilon_{ijk} x_j b_k \nonumber \\
&=&a_i+\sum_{jk} \epsilon_{ijk} \left(a_j+\sum_{lm}\epsilon_{jlm}x_l b_m\right) b_k \nonumber \\
&=&a_i+\sum_{jk} \epsilon_{ijk}a_j b_k+\sum_{jklm}(\delta_{im}\delta_{kl}-\delta_{il}\delta_{km}) x_l b_m b_k \nonumber \\
&=&a_i+(\mathbf{a}\times\mathbf{b})_i+(\mathbf{x}\cdot\mathbf{b})b_i-b^2 x_i \nonumber \\
&=&\frac{1}{1+b^2}\left[a_i+(\mathbf{a}\times\mathbf{b})_i+(\mathbf{a}\cdot\mathbf{b})b_i\right].
\end{eqnarray}
Here, we used $\sum_j \epsilon_{ijk}\epsilon_{mjl}=\delta_{im}\delta_{kl}-\delta_{il}\delta_{km}$ and $\mathbf{x}\cdot\mathbf{b}=(\mathbf{a}+\mathbf{x}\times\mathbf{b})\cdot\mathbf{b}=\mathbf{a}\cdot\mathbf{b}$. Thus, we have
\begin{equation}
\label{eq:self_consistent_form_x-solution}
\mathbf{x}=\frac{1}{1+b^2}\left[\mathbf{a}+(\mathbf{a}\times\mathbf{b})+(\mathbf{a}\cdot\mathbf{b})\mathbf{b}\right].
\end{equation}

Using the result of Eq.~\eqref{eq:self_consistent_form_x-solution},
we therefore obtain $\mathbf{G}$ as the following form: 
\begin{equation}
\begin{aligned}\label{AP7b}
\mathbf{G}=q\tau\Big(-\frac{\partial f^{(0)}_\mathbf{k}}{\partial \epsilon_\mathbf{k}}\Big)\frac{\mathbf{E}+\frac{\mu}{c}\mathbf{E}\times\mathbf{B}+\frac{\mu^2}{c^2}(\mathbf{E}\cdot\mathbf{B})\mathbf{B}}{1+\frac{\mu^2}{c^2}|\mathbf{B}|^2}.
\end{aligned}
\end{equation}
Note that Eq.~\eqref{AP7} is reduced to Eq.~\eqref{AP7b} when the mobility tensor is given by $\tilde{\mu}_{ij}=\mu\delta_{ij}$.

Finally, the nonequilibrium distribution function $g_\mathbf{k}=\tilde{\mathbf{v}}_{\mathbf{k}}\cdot\mathbf{G}$ is given by
\begin{equation}
\begin{aligned}
g=q\tau\Big(-\frac{\partial f^{(0)}_\mathbf{k}}{\partial \epsilon_\mathbf{k}}\Big)\tilde{\mathbf{v}}_\mathbf{k}\cdot\frac{\mathbf{E}+\frac{\mu}{c}\mathbf{E}\times\mathbf{B}+\frac{\mu^2}{c^2}(\mathbf{E}\cdot\mathbf{B})\mathbf{B}}{1+\frac{\mu^2}{c^2}|\mathbf{B}|^2}.
\end{aligned}
\end{equation}

\section{Magnetoconductivity of Weyl semimetals}
In this section, we derive the magnetoconductivity of Weyl semimetals including the Taylor expanded $D_{\mathbf k}^{-1}$. Here we set the magnetic field orientation as $\mathbf{B}=B_x\hat{x}+B_y\hat{y}$.

\subsection{Longitudinal magnetoconductivity}\label{B1}
Here we go through a detailed derivation of the longitudinal magnetoconductivity. Let us start with Eq.~\eqref{eqsimplesum} in the main text:
\begin{equation}\label{eqB1}
\sigma_{xx}=q^2\int\frac{d^3k}{(2\pi)^3}\frac{\tau(-\frac{\partial f^{(0)}_\mathbf{k}}{\partial \epsilon_\mathbf{k}})D^{-1}_\mathbf{k}}{1+\frac{\mu^2}{c^2}|\mathbf{B}|^2}\Big[\Sigma_{\mathbf k}^{(0)}+\Sigma_{\mathbf k}^{(1)}+\Sigma_{\mathbf k}^{(2)}\Big]\equiv \sigma^{(0)}_{xx}+\sigma^{(1)}_{xx}+\sigma^{(2)}_{xx},
\end{equation}
where $\Sigma^{(i)}_\mathbf{k}$ is a sum of the terms that include $i$th order of $\mu$ described in Eq.~\eqref{eqsimplesum-Sigma} in the main text which corresponds to conductivity of $\sigma^{(i)}_{xx}$ respectively. Inserting the Taylor expanded density of states correction
\begin{equation}
D^{-1}_\mathbf{k}=1+\frac{q}{\hbar c}(\mathbf{B}\cdot\mathbf{\Omega}_\mathbf{k})+(\frac{q}{\hbar c})^2(\mathbf{B}\cdot\mathbf{\Omega}_\mathbf{k})^2+\cdots
\end{equation}
to Eq.~\eqref{eqB1} and focusing on $\Sigma^{(0)}_\mathbf{k}$, $\sigma^{(0)}_{xx}$ yields
\begin{equation}
\begin{aligned}
\sigma^{(0)}_{xx}=q^2\!\!\int\frac{d^3k}{(2\pi)^3}\frac{\tau \delta(k-k_{\rm F})}{\hbar v_{\rm F}}\frac{1+\frac{q}{\hbar c}(\mathbf{B}\cdot\mathbf{\Omega}_\mathbf{k})+(\frac{q}{\hbar c})^2(\mathbf{B}\cdot\mathbf{\Omega}_\mathbf{k})^2}{1+\frac{\mu^2}{c^2}|\mathbf{B}|^2}\Big[v^2_x-\frac{2q}{\hbar c}v_x(\mathbf{v}_{\mathbf{k}}\cdot\mathbf{\Omega}_\mathbf{k})|\mathbf{B}|\cos\Gamma+\frac{q^2}{(\hbar c)^2}(\mathbf{v}_{\mathbf{k}}\cdot\mathbf{\Omega}_\mathbf{k})^2|\mathbf{B}|^2\cos^2\Gamma\Big],
\end{aligned}
\end{equation}
where we replaced $(-\frac{\partial f^{(0)}_\mathbf{k}}{\partial \epsilon_\mathbf{k}})=\frac{\delta(k-k_{\rm F})}{\hbar v_{\rm F}}$ for zero temperature. Expanding and throwing away terms that will give zero contribution after the momentum integral due to odd order in $k_i$ $(i=x,y,z)$, we are left with the following angular integral after switching to spherical coordinates then integrating $k$ out:
\begin{equation}
\sigma^\text{(0)}_{xx}=q^2\int^{2\pi}_0d\phi\int^\pi_0d\theta \frac{k_{\rm F}^2\sin\theta}{(2\pi)^3}\frac{\tau I^{(0)}(\theta,\phi)}{\hbar v_{\rm F}(1+b^2_\mu)},
\end{equation}
where 
\begin{equation}
\begin{aligned}
I^{(0)}(\theta,\phi)=v_{\rm F}^2\Big[&\sin^2\theta\cos^2\phi+b^2_\text{BC}\cos^2\Gamma-2b^2_\text{BC}\cos^2\Gamma \sin^2\theta \cos^2\phi\\
+&b^2_\text{BC}(\cos^2\Gamma \sin^4\theta \cos^4\phi+\sin^2\Gamma \sin^4\theta \cos^2\phi \sin^2\phi)\\
+&b^4_\text{BC}(\cos^4\Gamma \sin^2\theta \cos^2\phi +\cos^2\Gamma \sin^2\Gamma \sin^2\theta \sin^2\phi)\Big].
\end{aligned}
\end{equation}
Finally, carrying out the angular integral leads to 
\begin{equation}
\sigma^\text{(0)}_{xx}=\frac{\sigma_0}{1+b^2_\mu}\left[1+\frac{b^2_\text{BC}}{5}\left(8\cos^2\Gamma+\sin^2\Gamma\right)+b^4_\text{BC}\cos^2\Gamma\right],
\end{equation}
where $\sigma_0=q^2(\frac{k_{\rm F}^2}{2\pi^2\hbar v_{\rm F}})(\frac{\tau v_{\rm F}^2}{3})=q^2N_0D$ is the longitudinal magnetoconductivity in the absence of magnetic field.
  
Focusing now on $\Sigma^{(1)}_\mathbf{k}$, $\sigma^{(1)}_{xx}$ yields
\begin{equation}
\sigma^{(1)}_{xx}=q^2\!\!\int \frac{d^3k}{(2\pi)^3}\frac{\tau \delta(k-k_{\rm F})}{\hbar v_{\rm F}}\frac{1+\frac{q}{\hbar c}(\mathbf{B}\cdot\mathbf{\Omega}_\mathbf{k})+(\frac{q}{\hbar c})^2(\mathbf{B}\cdot\mathbf{\Omega}_\mathbf{k})^2}{1+\frac{\mu^2}{c^2}|\mathbf{B}|^2}\frac{\mu}{c}\Big[-v_x(\mathbf{v}_\mathbf{k}\times\mathbf{B})_x+\frac{q}{\hbar c}(\mathbf{v}_\mathbf{k}\times\mathbf{B})_x(\mathbf{v}_\mathbf{k}\cdot\mathbf{\Omega}_\mathbf{k})|\mathbf{B}|\cos\Gamma\Big].
\end{equation}
The above whole expression vanishes after the momentum integral, because  $(\mathbf{v}_\mathbf{k}\times\mathbf{B})_x=v_yB_z-v_zB_y=-v_zB_y$ as $B_z=0$. This will lead to a single order in $v_z$ in every term in the integrand therefore result in zero after integration. 

Finally, focusing on $\Sigma^{(2)}_\mathbf{k}$, $\sigma^{(2)}_{xx}$ yields
\begin{eqnarray}
\sigma^{(2)}_{xx}=q^2\!\!\int \frac{d^3k}{(2\pi)^3}\!\!\!\!\!&&\frac{\tau \delta(k-k_{\rm F})}{\hbar v_{\rm F}}\frac{1+\frac{q}{\hbar c}(\mathbf{B}\cdot\mathbf{\Omega}_\mathbf{k})+(\frac{q}{\hbar c})^2(\mathbf{B}\cdot\mathbf{\Omega}_\mathbf{k})^2}{1+\frac{\mu^2}{c^2}|\mathbf{B}|^2}\Big[(\mathbf{v}_\mathbf{k}\cdot\mathbf{B})v_x|\mathbf{B}|\cos\Gamma-\frac{q}{\hbar c}(\mathbf{v}_\mathbf{k}\cdot\mathbf{\Omega}_\mathbf{k})|\mathbf{B}|^3v_x\cos\Gamma \nonumber \\
&-&\frac{q}{\hbar c}(\mathbf{v}_\mathbf{k}\cdot\mathbf{B})(\mathbf{v}_\mathbf{k}\cdot\mathbf{\Omega}_\mathbf{k})|\mathbf{B}|^2\cos^2\Gamma
-\frac{q^2}{(\hbar c)^2}(\mathbf{v}_\mathbf{k}\cdot\mathbf{\Omega}_\mathbf{k})^2|\mathbf{B}|^4\cos^2\Gamma\Big].
\end{eqnarray}
Simplifying the product of the Taylor expanded $D^{-1}_\mathbf{k}$ and terms in the square bracket while again, keeping only the non-zero contribution, we are left with the following angular integral after switching to spherical coordinates then integrating $k$ out:
\begin{equation}
\sigma^{(2)}_{xx}=q^2\int^{2\pi}_0d\phi\int^\pi_0d\theta\frac{k_{\rm F}^2\sin\theta}{(2\pi)^3}\frac{\tau I^{(2)}(\theta,\phi)}{\hbar v_{\rm F}(1+b^2_\mu)},
\end{equation}
where
\begin{equation}
\begin{aligned}
I^{(2)}(\theta,\phi)=v^2_{\rm F}b^2_\mu
\Big[&\sin^2\theta\cos^2\phi\cos^2\Gamma+b^2_\text{BC}\big(\cos^2\Gamma-\cos^2\Gamma\sin^2\theta\cos^2\phi-\sin^2\theta\cos^2\phi\cos^4\Gamma\\
-&\sin^2\theta\sin^2\phi\sin^2\Gamma\cos^2\Gamma+\sin^4\theta\cos^4\phi\cos^4\Gamma+3\sin^4\theta\cos^2\phi\sin^2\phi\cos^2\Gamma\sin^2\Gamma\big)\\
+&b^4_\text{BC}\left(\sin^2\theta\cos^2\phi\cos^4\Gamma+\sin^2\theta\sin^2\phi\sin^2\Gamma\cos^2\Gamma\right)\Big].
\end{aligned}
\end{equation}
Finally, carrying out the angular integral leads to
\begin{equation}
\sigma^{(2)}_{xx}=\sigma_0\frac{b^2_\mu}{1+b^2_\mu}\left(\cos^2\Gamma+\frac{8}{5}b^2_\text{BC}\cos^2\Gamma+b^4_\text{BC}\cos^2\Gamma\right).
\end{equation}

Adding up the $\sigma^{(i)}_{xx}$s, we have
\begin{equation}
\begin{aligned}
\sigma_{xx}=&\frac{\sigma_0}{1+b^2_\mu}\left[1+\frac{b^2_\text{BC}}{5}\left(8\cos^2\Gamma+\sin^2\Gamma\right)+b^4_\text{BC}\cos^2\Gamma+b^2_\mu\left(\cos^2\Gamma+\frac{8}{5}b^2_\text{BC}\cos^2\Gamma+b^4_\text{BC}\cos^2\Gamma\right)\right].
\end{aligned}
\end{equation}

\subsection{Planar Hall conductivity}\label{B2}
Here we go through a detailed derivation of the planar Hall conductivity. Starting with Eq.~(20) in the main text for $d=3$, we again express it in powers of $\mu$ while using the definition for $\tilde{\mathbf{v}}_\mathbf{k}$. We then obtain,
\begin{equation}\label{planar hall cond}
\sigma_{yx}=q^2\int\frac{d^3k}{(2\pi)^3}\frac{\tau(-\frac{\partial f^{(0)}_\mathbf{k}}{\partial \epsilon_\mathbf{k}})D^{-1}_\mathbf{k}}{1+\frac{\mu^2}{c^2}|\mathbf{B}|^2}\Big[\Sigma_{\mathbf k}^{(0)}+\Sigma_{\mathbf k}^{(1)}+\Sigma_{\mathbf k}^{(2)}\Big]\equiv \sigma^{(0)}_{yx}+\sigma^{(1)}_{yx}+\sigma^{(2)}_{yx},
\end{equation}
where
\begin{subequations}\label{eqsimplesum-Sigma1}
\begin{alignat}{4}
\Sigma_{\mathbf k}^{(0)}=&~v_yv_x-\frac{q}{\hbar c}v_y(\mathbf{v}_\mathbf{k}\cdot\mathbf{\Omega}_\mathbf{k})|\mathbf{B}|\cos\Gamma-\frac{q}{\hbar c}v_x(\mathbf{v}_\mathbf{k}\cdot\mathbf{\Omega}_\mathbf{k})|\mathbf{B}|\sin\Gamma+\left(\frac{q}{\hbar c}\right)^2(\mathbf{v}_\mathbf{k}\cdot\mathbf{\Omega}_\mathbf{k})^2|\mathbf{B}|^2\cos\Gamma\sin\Gamma,\\
\nonumber\\
\Sigma_{\mathbf k}^{(1)}=&~\frac{\mu}{c}\left[-v_y(\mathbf{v}_\mathbf{k}\times\mathbf{B})_x+\frac{q}{\hbar c}(\mathbf{v}_\mathbf{k}\cdot\mathbf{\Omega}_\mathbf{k})(\mathbf{v}_\mathbf{k}\times\mathbf{B})_x|\mathbf{B}|\sin\Gamma\right],\\
\nonumber\\
\Sigma_{\mathbf k}^{(2)}=&~\frac{\mu^2}{c^2}\Big[(\mathbf{v}_\mathbf{k}\cdot\mathbf{B})v_y|\mathbf{B}|\cos\Gamma-\frac{q}{\hbar c}(\mathbf{v}_\mathbf{k}\cdot\mathbf{\Omega}_\mathbf{k})|\mathbf{B}|^3v_y\cos\Gamma-\frac{q}{\hbar c}(\mathbf{v}_\mathbf{k}\cdot\mathbf{B})(\mathbf{v}_\mathbf{k}\cdot\mathbf{\Omega}_\mathbf{k})|\mathbf{B}|^2\cos\Gamma\sin\Gamma\nonumber\\
+&\left(\frac{q}{\hbar c}\right)^2(\mathbf{v}_\mathbf{k}\cdot\mathbf{\Omega}_\mathbf{k})^2|\mathbf{B}|^4\cos\Gamma\sin\Gamma\Big],
\end{alignat}
\end{subequations}
each corresponds to planar Hall conductivity of $\sigma^{(i)}_{yx}$ respectively. Inserting the Taylor expanded density of states correction
\begin{equation}
D^{-1}_\mathbf{k}=1+\frac{q}{\hbar c}(\mathbf{B}\cdot\mathbf{\Omega}_\mathbf{k})+(\frac{q}{\hbar c})^2(\mathbf{B}\cdot\mathbf{\Omega}_\mathbf{k})^2+\cdots
\end{equation}
to Eq.~\eqref{planar hall cond} and focusing on $\Sigma^{(0)}_\mathbf{k}$, $\sigma^{(0)}_{yx}$ yields
\begin{eqnarray}
\sigma^{(0)}_{yx}=q^2\int\frac{d^3k}{(2\pi)^3}\frac{\tau \delta(k-k_{\rm F})}{\hbar v_{\rm F}}\frac{1+\frac{q}{\hbar c}(\mathbf{B}\cdot\mathbf{\Omega}_\mathbf{k})+(\frac{q}{\hbar c})^2(\mathbf{B}\cdot\mathbf{\Omega}_\mathbf{k})^2}{1+\frac{\mu^2}{c^2}|\mathbf{B}|^2}&&\!\!\!\!\!\!\Big[v_yv_x-\frac{q}{\hbar c}v_y(\mathbf{v}_\mathbf{k}\cdot\mathbf{\Omega}_\mathbf{k})|\mathbf{B}|\cos\Gamma-\frac{q}{\hbar c}v_x(\mathbf{v}_\mathbf{k}\cdot\mathbf{\Omega}_\mathbf{k})|\mathbf{B}|\sin\Gamma \nonumber\\
&+&\left(\frac{q}{\hbar c}\right)^2(\mathbf{v}_\mathbf{k}\cdot\mathbf{\Omega}_\mathbf{k})^2|\mathbf{B}|^2\cos\Gamma\sin\Gamma\Big],
\end{eqnarray}
where we replaced $(-\frac{\partial f^{(0)}_\mathbf{k}}{\partial \epsilon_\mathbf{k}})=\frac{\delta(k-k_{\rm F})}{\hbar v_{\rm F}}$ for zero temperature. Expanding and throwing away terms that will give zero contribution after the momentum integral due to odd order in $k_i$ $(i=x,y,z)$, we are left with the following angular integral after switching to spherical coordinates then integrating $k$ out:
\begin{equation}
\sigma^{(0)}_{yx}=q^2\int^{2\pi}_0d\phi\int^\pi_0d\theta\frac{k_{\rm F}^2\sin\theta}{(2\pi)^3}\frac{\tau I^{(0)}(\theta,\phi)}{\hbar v_{\rm F}(1+b^2_\mu)},
\end{equation}
where
\begin{equation}
\begin{aligned}
I^{(0)}(\theta,\phi)=v_{\rm F}^2\Big[&b^2_\text{BC}\big(\cos\Gamma\sin\Gamma-\sin^2\theta\cos^2\phi\cos\Gamma\sin\Gamma\\
-&\sin^2\theta\sin^2\phi\cos\Gamma\sin\Gamma+2\sin^4\theta\cos^2\phi\sin^2\phi\cos\Gamma\sin\Gamma\big)\\
+&b^4_\text{BC}\big(\sin^2\theta\cos^2\phi\cos^3\Gamma\sin\Gamma+\sin^2\theta\sin^2\phi\cos\Gamma\sin^3\Gamma\big)\Big].
\end{aligned}
\end{equation}
Finally, carrying out the angular integral leads to 
\begin{equation}
\sigma^{(0)}_{yx}=\frac{\sigma_0}{1+b^2_\mu}\left(\frac{7}{5}b^2_\text{BC}+b^4_\text{BC}\right)\cos\Gamma\sin\Gamma.
\end{equation}

Focusing now on $\Sigma^{(1)}_\mathbf{k}$, $\sigma^{(1)}_{yx}$ yields
\begin{equation}
\sigma^{(1)}_{yx}=q^2\int\frac{d^3k}{(2\pi)^3}\frac{\tau \delta(k-k_{\rm F})}{\hbar v_{\rm F}}\frac{1+\frac{q}{\hbar c}(\mathbf{B}\cdot\mathbf{\Omega}_\mathbf{k})+(\frac{q}{\hbar c})^2(\mathbf{B}\cdot\mathbf{\Omega}_\mathbf{k})^2}{1+\frac{\mu^2}{c^2}|\mathbf{B}|^2}\frac{\mu}{c}\left[-v_y(\mathbf{v}_\mathbf{k}\times\mathbf{B})_x+\frac{q}{\hbar c}(\mathbf{v}_\mathbf{k}\cdot\mathbf{\Omega}_\mathbf{k})(\mathbf{v}_\mathbf{k}\times\mathbf{B})_x|\mathbf{B}|\sin\Gamma\right].
\end{equation}
The above whole expression vanishes after the momentum integral, because  $(\mathbf{v}_\mathbf{k}\times\mathbf{B})_x=v_yB_z-v_zB_y=-v_zB_y$ as $B_z=0$. This will lead to a single order in $v_z$ in every term in the integrand therefore result in zero after integration.\\

Finally, focusing on $\Sigma^{(2)}_\mathbf{k}$, $\sigma^{(2)}_{yx}$ yields
\begin{equation}
\begin{aligned}
\sigma^{(2)}_{yx}=q^2\int\frac{d^3k}{(2\pi)^3}\frac{\tau \delta(k-k_{\rm F})}{\hbar v_{\rm F}}\frac{1+\frac{q}{\hbar c}(\mathbf{B}\cdot\mathbf{\Omega}_\mathbf{k})+(\frac{q}{\hbar c})^2(\mathbf{B}\cdot\mathbf{\Omega}_\mathbf{k})^2}{1+\frac{\mu^2}{c^2}|\mathbf{B}|^2}\frac{\mu^2}{c^2}\Big[(\mathbf{v}_\mathbf{k}\cdot\mathbf{B})v_y|\mathbf{B}|\cos\Gamma-\frac{q}{\hbar c}(\mathbf{v}_\mathbf{k}\cdot\mathbf{\Omega}_\mathbf{k})|\mathbf{B}|^3v_y\cos\Gamma\\
-\frac{q}{\hbar c}(\mathbf{v}_\mathbf{k}\cdot\mathbf{B})(\mathbf{v}_\mathbf{k}\cdot\mathbf{\Omega}_\mathbf{k})|\mathbf{B}|^2\cos\Gamma\sin\Gamma+\left(\frac{q}{\hbar c}\right)^2(\mathbf{v}_\mathbf{k}\cdot\mathbf{\Omega}_\mathbf{k})^2|\mathbf{B}|^4\cos\Gamma\sin\Gamma\Big].
\end{aligned}
\end{equation}
Simplifying the product of the Taylor expanded $D^{-1}_\mathbf{k}$ and terms in the square bracket while again, keeping only the non-zero contribution, we are left with the following angular integral after switching to spherical coordinates then integrating $k$ out:
\begin{equation}
\sigma^{(2)}_{yx}=q^2\int^{2\pi}_0d\phi\int^\pi_0d\theta\frac{k_{\rm F}^2\sin\theta}{(2\pi)^3}\frac{\tau I^{(2)}(\theta,\phi)}{\hbar v_{\rm F}(1+b^{-2}_\mu)},
\end{equation}
where
\begin{eqnarray}
I^{(2)}(\theta,\phi)=v_{\rm F}^2 b^2_{\mu}&\Big[&\sin^3\theta\sin^2\phi\cos\Gamma\sin\Gamma+b^2_\text{BC}\big(\sin\theta\cos\Gamma\sin\Gamma-\sin^3\theta\cos^2\phi\cos^3\Gamma\sin\Gamma-\sin^3\theta\sin^2\phi\sin\Gamma\cos\Gamma\nonumber\\&-&\sin^3\theta\sin^2\phi\cos\Gamma\sin^3\Gamma+\sin^5\theta\cos^2\phi\sin^2\phi\cos^3\Gamma\sin\Gamma+2\sin^5\theta\cos^2\phi\sin^2\phi\cos^3\Gamma\sin\Gamma\nonumber\\&+&\sin^5\theta\sin^4\phi\cos\Gamma\sin^3\Gamma\big)+b^4_\text{BC}\big(\sin^3\theta\cos^2\phi\cos^3\Gamma\sin\Gamma+\sin^3\theta\sin^2\phi\cos\Gamma\sin^3\Gamma\big)\Big].
\end{eqnarray}
Finally, carrying out the angular integral leads to 
\begin{equation}
\sigma^{(2)}_{yx}=\sigma_0\frac{b^2_\mu}{1+b^2_\mu}\Big(1+\frac{8}{5}b^2_\text{BC}+b^4_\text{BC}\Big)\cos\Gamma\sin\Gamma.
\end{equation}
Adding up the $\sigma^{(i)}_{yx}$s, we have
\begin{equation}
\sigma_{yx}=\frac{\sigma_0}{1+b^2_\mu}\left[\left(\frac{7}{5}b^2_\text{BC}+b^4_\text{BC}\right)\cos\Gamma\sin\Gamma+b^2_\mu\Big(1+\frac{8}{5}b^2_\text{BC}+b^4_\text{BC}\Big)\cos\Gamma\sin\Gamma\right].
\end{equation}

\subsection{Conductivity $\sigma_{zx}$}\label{B3}
Here we go through a detailed derivation of $\sigma_{zx}$. Starting with Eq.~(20) in the main text for $d=3$, we again express it in powers of $\mu$ while using the definition for $\tilde{v}_\mathbf{k}$. We then obtain
\begin{equation}\label{hall cond}
\sigma_{zx}=q^2\int\frac{d^3k}{(2\pi)^3}\frac{\tau(-\frac{\partial f^{(0)}_\mathbf{k}}{\partial \epsilon_\mathbf{k}})D^{-1}_\mathbf{k}}{1+\frac{\mu^2}{c^2}|\mathbf{B}|^2}\Big[\Sigma_{\mathbf k}^{(0)}+\Sigma_{\mathbf k}^{(1)}+\Sigma_{\mathbf k}^{(2)}\Big]\equiv \sigma^{(0)}_{zx}+\sigma^{(1)}_{zx}+\sigma^{(2)}_{zx},
\end{equation}
where
\begin{subequations}\label{eqsimplesum-Sigma2}
\begin{alignat}{4}
\Sigma_{\mathbf k}^{(0)}=&v_zv_x-\frac{q}{\hbar c}(\mathbf{v}_\mathbf{k}\cdot\mathbf{\Omega}_\mathbf{k})v_z|\mathbf{B}|\cos\Gamma,\\
\nonumber\\
\Sigma_{\mathbf k}^{(1)}=&~-\frac{\mu}{c}v_z\left(\mathbf{v}_\mathbf{k}\times\mathbf{B}\right)_x,\\
\nonumber\\
\Sigma_{\mathbf k}^{(2)}=&~\frac{\mu^2}{c^2}\left[(\mathbf{v}_\mathbf{k}\cdot\mathbf{B})-\frac{q}{\hbar c}(\mathbf{v}_\mathbf{k}\cdot\mathbf{\Omega}_\mathbf{k})|\mathbf{B}|^2\right]v_z|\mathbf{B}|\cos\Gamma,
\end{alignat}
\end{subequations}
each corresponds to Hall conductivity of $\sigma^{(i)}_{zx}$ respectively. Inserting the Taylor expanded density of states correction to Eq.~\eqref{hall cond} and focusing on $\Sigma^{(0)}_\mathbf{k}$, $\sigma^{(0)}_{zx}$ yields
\begin{equation}\label{eq:sigma_zx_0}
\begin{aligned}
\sigma^{(0)}_{zx}=q^2\int\frac{d^3k}{(2\pi)^3}\frac{\tau \delta(k-k_{\rm F})}{\hbar v_{\rm F}}\frac{1+\frac{q}{\hbar c}(\mathbf{B}\cdot\mathbf{\Omega}_\mathbf{k})+(\frac{q}{\hbar c})^2(\mathbf{B}\cdot\mathbf{\Omega}_\mathbf{k})^2}{1+\frac{\mu^2}{c^2}|\mathbf{B}|^2}\Big[v_zv_x-\frac{q}{\hbar c}(\mathbf{v}_\mathbf{k}\cdot\mathbf{\Omega}_\mathbf{k})v_z|\mathbf{B}|\cos\Gamma\Big],
\end{aligned}
\end{equation}
where we replaced $(-\frac{\partial f^{(0)}_\mathbf{k}}{\partial \epsilon_\mathbf{k}})=\frac{\delta(k-k_{\rm F})}{\hbar v_{\rm F}}$ for zero temperature. The above whole expression in Eq.~\eqref{eq:sigma_zx_0} vanishes after the momentum integral.

Focusing now on $\Sigma^{(1)}_\mathbf{k}$, $\sigma^{(1)}_{zx}$ yields
\begin{equation}
\begin{aligned}
\sigma^{(1)}_{zx}=q^2\int\frac{d^3k}{(2\pi)^3}\frac{\tau \delta(k-k_{\rm F})}{\hbar v_{\rm F}}\frac{1+\frac{q}{\hbar c}(\mathbf{B}\cdot\mathbf{\Omega}_\mathbf{k})+(\frac{q}{\hbar c})^2(\mathbf{B}\cdot\mathbf{\Omega}_\mathbf{k})^2}{1+\frac{\mu^2}{c^2}|\mathbf{B}|^2}\Big[-\frac{\mu}{c}v_z\left(\mathbf{v}_\mathbf{k}\times\mathbf{B}\right)_x\Big].
\end{aligned}
\end{equation}
Expanding and throwing away terms that will give zero contribution after the momentum integral due to odd order in $k_i$ $(i=x,y,z)$, we are left with the following angular integral after switching to spherical coordinates then integrating $k$ out:
\begin{equation}
\sigma^{(2)}_{zx}=q^2\int^{2\pi}_0d\phi\int^\pi_0d\theta \frac{k_{\rm F}^2\sin\theta}{(2\pi)^3}\frac{\tau b_\mu I^{(1)}(\theta,\phi)}{\hbar v_{\rm F}(1+b^2_{\mu})},
\end{equation}
where 
\begin{equation}
I^{(1)}(\theta,\phi)=v_{\rm F}^2\Big[\cos^2\theta\sin\theta\sin\Gamma+b^2_\text{BC}\big(\cos^2\theta\sin^3\theta\cos^2\phi\cos^2\Gamma\sin\Gamma+\cos^2\theta\sin^3\theta\sin^2\phi\sin^3\Gamma\big)\Big].
\end{equation}
Finally, carrying out the angular integral leads to 
\begin{equation}
\sigma^{(1)}_{zx}=\sigma_0\frac{b_\mu}{1+b^2_\mu}\left(1+\frac{1}{5}b^2_\text{BC}\right)\sin\Gamma.
\end{equation}

Finally, focusing on $\Sigma^{(2)}_\mathbf{k}$, $\sigma^{(2)}_{zx}$ yields
\begin{equation}
\sigma^{(0)}_{zx}=q^2\int\frac{d^3k}{(2\pi)^3}\frac{\tau \delta(k-k_{\rm F})}{\hbar v_{\rm F}}\frac{1+\frac{q}{\hbar c}(\mathbf{B}\cdot\mathbf{\Omega}_\mathbf{k})+(\frac{q}{\hbar c})^2(\mathbf{B}\cdot\mathbf{\Omega}_\mathbf{k})^2}{1+\frac{\mu^2}{c^2}|\mathbf{B}|^2}\frac{\mu^2}{c^2}\left[(\mathbf{v}_\mathbf{k}\cdot\mathbf{B})-\frac{q}{\hbar c}(\mathbf{v}_\mathbf{k}\cdot\mathbf{\Omega}_\mathbf{k})|\mathbf{B}|^2\right]v_z|\mathbf{B}|\cos\Gamma.
\end{equation}
The above expression vanishes after the momentum integral.

Adding up the $\sigma^{(i)}_{zx}$s, we have
\begin{equation}\label{eq:sigma_zx}
\sigma_{zx}=\sigma_0\frac{b_\mu}{1+b^2_\mu}\left(1+\frac{1}{5}b^2_\text{BC}\right)\sin\Gamma.
\end{equation} 
Note that $\sigma_{zx}$ at $\Gamma={\pi\over 2}$ in Eq.~\eqref{eq:sigma_zx} corresponds to the Hall conductivity.

\subsection{Conductivity $\sigma_{zz}$}
Here we go through a detailed derivation of the $\sigma_{zz}$. Starting with Eq.~(20) in the main text for $d=3$, we again express it in powers of $\mu$ while using the definition for $\tilde{\mathbf{v}}_\mathbf{k}$. We then obtain
\begin{equation}\label{sigma zz}
\sigma_{zz}=q^2\int\frac{d^3k}{(2\pi)^3}\frac{\tau(-\frac{\partial f^{(0)}_\mathbf{k}}{\partial \epsilon_\mathbf{k}})D^{-1}_\mathbf{k}}{1+\frac{\mu^2}{c^2}|\mathbf{B}|^2}\Big[\Sigma_{\mathbf k}^{(0)}+\Sigma_{\mathbf k}^{(1)}+\Sigma_{\mathbf k}^{(2)}\Big]\equiv \sigma^{(0)}_{zz}+\sigma^{(1)}_{zz}+\sigma^{(2)}_{zz},
\end{equation}
where
\begin{subequations}\label{eqsimplesum-Sigma3}
\begin{alignat}{4}
\Sigma_{\mathbf k}^{(0)}=&v^2_z,\\
\nonumber\\
\Sigma_{\mathbf k}^{(1)}=&~-\frac{\mu}{c}v_z\left(\mathbf{v}_\mathbf{k}\times\mathbf{B}\right)_z,\\
\nonumber\\
\Sigma_{\mathbf k}^{(2)}=&~0,
\end{alignat}
\end{subequations}
as $B_z=0$. Note that after the momentum integral, $\Sigma^{(1)}_\mathbf{k}$ will vanish due to odd order in $v_z$. The only remaining contribution is from $\Sigma^{(0)}_\mathbf{k}$ which gives
\begin{equation}
\begin{aligned}
\sigma^{(0)}_{zz}=&q^2\int\frac{d^3k}{(2\pi)^3}\frac{D^{-1}_\mathbf{k}\tau(-\frac{\partial f^{(0)}_\mathbf{k}}{\partial \epsilon_\mathbf{k}})}{1+\frac{\mu^2}{c^2}|\mathbf{B}|^2}v^2_z\\
=&q^2\int\frac{d\Omega}{\hbar v_{\rm F}(2\pi)^3}\frac{\tau v_{\rm F}^2k_{\rm F}^2}{1+b^2_\mu}\Big(\cos^2\theta+b^2_\text{BC}\cos^2\theta\sin^2\theta\cos^2\phi\cos^2\Gamma+b^2_\text{BC}\cos^2\theta\sin^2\theta\sin^2\phi\sin^2\Gamma\Big)\sin\theta\\
=&\frac{\sigma_0}{1+b^2_\mu}\left(1+\frac{1}{5}b^2_\text{BC}\right).
\end{aligned}
\end{equation}

\subsection{Conductivity and resistivity tensors}\label{B4}

Finally, we have the following conductivity tensor including the Taylor expanded $D_{\mathbf k}^{-1}$:
\begin{eqnarray}
{\bm{\sigma}(\mathbf{B})}&=&\frac{\sigma_0}{1+b^2_{\mu}}\\ &\times&
\begin{bmatrix} 1+\frac{1}{5}b^2_{\rm BC}+\Big[\frac{7}{5}b^2_{\rm BC}+b_{\mu}^2\Big(1+\frac{8}{5}b^2_{\rm BC}\Big)\Big]\cos^2{\Gamma}& \Big[{b^2_{\mu}}(1+\frac{1}{5}b^{2}_{\rm BC})+\frac{7}{5}(1+b^2_{\mu})b^{2}_{\rm BC}\Big]\sin{\Gamma}\cos{\Gamma} & -b_{\mu}\left(1+\frac{1}{5}b^2_\text{BC}\right)\sin\Gamma \\ \Big[{b^2_{\mu}}(1+\frac{1}{5}b^{2}_{\rm BC})+\frac{7}{5}(1+b^2_{\mu})b^{2}_{\rm BC}\Big]\sin{\Gamma}\cos{\Gamma} & 1+\frac{1}{5}b^2_{\rm BC}+\Big[\frac{7}{5}b^2_{\rm BC}+b_{\mu}^2\Big(1+\frac{8}{5}b^2_{\rm BC}\Big)\Big]\sin^2{\Gamma}& b_{\mu}\left(1+\frac{1}{5}b^2_\text{BC}\right)\cos\Gamma \\ b_{\mu}\left(1+\frac{1}{5}b^2_\text{BC}\right)\sin\Gamma & -b_{\mu}\left(1+\frac{1}{5}b^2_\text{BC}\right)\cos\Gamma & 1+\frac{1}{5}b^2_\text{BC} \nonumber\\ 
\end{bmatrix}.
\end{eqnarray}
The resistivity tensor $\bm{\rho}(\mathbf{B})$ is then obtained by the inverse of the conductivity tensor $\bm{\sigma}(\mathbf{B})$ as
\begin{eqnarray}
{\bm{\rho}(\mathbf{B})}&=&\rho_0
\begin{bmatrix}
\frac{50+10b^2_\text{BC}+70 b^2_\text{BC}\sin^2\Gamma}{50+90 b^2_\text{BC}+16 b^4_\text{BC}} & -\frac{35 b^2_\text{BC}\cos\Gamma \sin\Gamma}{25+45b^2_\text{BC}+8b^4_\text{BC}}&  \frac{5 b_\mu \sin\Gamma}{5+b^2_\text{BC}} \\
-\frac{35 b^2_\text{BC}\cos\Gamma \sin\Gamma}{25+45b^2_\text{BC}+8b^4_\text{BC}} & \frac{50+10b^2_\text{BC}+70 b^2_\text{BC}\cos^2\Gamma}{50+90 b^2_\text{BC}+16 b^4_\text{BC}}& -\frac{5 b_\mu \cos\Gamma}{5+b^2_\text{BC}}\\
-\frac{5 b_\mu \sin\Gamma}{5+b^2_\text{BC}} &\frac{5 b_\mu \cos\Gamma}{5+b^2_\text{BC}} &\frac{5}{5+b^2_\text{BC}} \\
\end{bmatrix}.
\end{eqnarray}
Note that although both the Lorentz force and the Berry curvature induced terms are present in $\sigma_{ij}(\mathbf{B})$, the Lorentz force induced terms do not appear in the resistivity $\rho_{ij}(\mathbf{B})$ where $i, j = x, y$ even if we include the Taylor expanded $D^{-1}_\mathbf{k}$.

\end{document}